\documentclass[aps,twocolumn]{revtex4}
\usepackage [dvips]{graphicx}
\begin{document}

\title{Thermalisation time and specific heat of neutron stars crust}

\author{
M. Fortin$^{a,b}$,
F. Grill$^{c}$, 
J. Margueron$^{b}$, 
N. Sandulescu$^{d}$\footnote{corresponding author 
(email:sandulescu@theory.nipne,ro)}
}
\affiliation {
$^{a}$ \'Ecole Normale Sup\'erieure, Departement de Physique, 24 rue Lhomond, 75005 Paris, France \\
$^{b}$ Institut de Physique Nucl\'eaire, Universit\'e Paris-Sud, F-91406 Orsay CEDEX, 
France \\
$^{c}$ Dipartimento di Fisica, Universit\'a degli Studi di Milano, 
Via Celoria 16, 20133 Milan, Italy \\ 
$^{d}$ National Institute of Physics and Nuclear Engineering, 76900,
Bucharest, Romania}

\begin{abstract}
We discuss the thermalisation process of the neutron stars crust described 
by solving the heat transport equation with a microscopic input for the specific heat
of baryonic matter. The heat equation is solved with initial conditions 
specific to a rapid cooling of the core. To calculate the specific heat of inner crust
baryonic matter, i.e., nuclear clusters and unbound neutrons, we use the quasiparticle 
spectrum provided by the Hartree-Fock-Bogoliubov approach at finite temperature. 
In this framework we analyse the dependence of the crust thermalisation on pairing 
properties and on cluster structure of inner crust matter. It is shown that the 
pairing correlations reduce the crust thermalisation time by a very large fraction. 
The calculations show also that the nuclear clusters have a non-negligible influence
on the time evolution of the surface temperature of the neutron star.
\end{abstract}

\maketitle

\section{Introduction}
The thermalisation process of the neutron stars crust can give important informations
about the properties of the crust matter. This is indeed the case in the rapid
cooling of isolated neutron stars \cite{lattimer} and in the thermal
afterburst relaxation of neutron stars from X-ray transients \cite{potekhin}.
In the rapid cooling process an important quantity is the cooling or thermalisation time 
of the crust, defined as the time needed for the crust matter to arrive to the temperature of the
cold core. Since the outer crust and the envelope have high thermal conductivity, the cooling 
time is essentially determined  by the inner crust matter, formed by nuclear clusters, unbound 
neutrons and ultrarelativistic electrons. 

The thermalisation time of the crust depends essentially on the
crust thickness \cite{lattimer}. However, several studies have shown
that the thermalisation time depends also significantly on the superfluid 
properties of the inner crust baryonic matter \cite{lattimer,pizzochero,monrozeau,ns}.
This dependence is induced
through the specific heat of unbound neutrons, strongly affected by the pairing energy gap. 
Since the neutron pairing gap is influenced by the presence of the nuclear clusters 
\cite{barranco,pizzochero,sandulescu1}, a reliable calculation 
of neutron specific heat should take the clusters into account.  How  the  intensity of pairing 
correlations affects the specific heat of the neutrons in the presence of nuclear clusters was analysed 
in Ref. \cite{sandulescu2}. Thus, using the framework of Hartree-Fock-Bogoliubov (HFB) approach 
at finite temperature, it was shown that  the specific heat can change with several orders of magnitude if the 
pairing gap is adjusted to describe two possible scenarios for neutron matter superfluidity, 
i.e., one corresponding to BCS approximation and the other to calculations schemes which 
take into account in-medium effects \cite{lombardo}. The impact which these changes in the specific 
heat could have on the thermalisation time was discussed in Ref. \cite{monrozeau}. Employing
a simple random walk model for the  cooling \cite{brown,pizzochero}, in which the diffusion 
of the heat towards the core was calculated without taking into account the dynamical change of the 
temperature through the whole crust, it was shown that the thermalisation times corresponding  to the two 
pairing scenarios mentioned above differs by a large fraction, which could be easily 
discriminated observationally.  The scope of this paper is to present more accurate estimations 
of the thermalisation time obtained by employing a more realistic cooling model based on dynamical 
solutions of the heat equations and on a state-of-the-art description of the specific heat
for baryonic matter.

\section{The model of crust thermalisation}
 
The crust thermalisation is described here in the rapid cooling scenario
in which the core arrives quickly to a much smaller temperature than the crust.
Due to this temperature inversion the heat stored in the crust diffuses into 
the core where it is dissipated by the neutrinos. The heat diffusion through the crust
can be described by the relativistic heat equation \cite{heat}:
\begin{equation} 
\frac{\partial}{\partial r} [ \frac{K r^2}{\Gamma(r)} e^\phi 
\frac{\partial }{\partial r} (e^\phi T) ] = 
r^2 \Gamma(r) e^\phi ( C_V \frac{\partial T}{\partial t}
+ e^{\phi}Q_{\nu} ),
\end{equation}
where T is the temperature, $t$ is the time,
$K$ is the thermal conductivity, $C_V$ is the specific heat and
$Q_\nu$ is the neutrino emissivity. The effect of the gravity is 
given through the  gravitational potential $\phi$, which enters in the
definition of the readshifted temperature $\tilde{T} = Te^{\phi}$, 
and the quantity
$\Gamma(r)=\left( 1-2Gm(r)/rc^2 \right)^{-1/2}$, where  $G$ is the 
gravitational constant and $m(r)$ is the gravitational mass included 
in a sphere of radius $r$. The latter is obtained solving the 
Tolman-Oppenheimer-Volkoff equation. 

In principle, the radial mass distribution and the properties of the inner
crust employed in the heat equation should be determined consistently from
the same equation of state (EOS) of nuclear matter. To describe in an
unified way both the crust and the core here we use the EOS given by the 
Skyrme interaction SLy4 \cite{sly4}, which provides a reasonable star 
model \cite{douchin} for the present study. The cooling calculations are done for 
a neutron star with a mass equal to 1.44 $M_\odot$. For this mass it is predicted
a total star radius of 11.59~ km and a central density of 3.51 $\rho_{0}$, 
where $\rho_{0}= 2.9\times 10^{14}$ g.cm$^{-3}$  is the saturation density for 
symmetric nuclear matter. The inner crust, defined here as the 
part of the star with the density ranging between $\rho_{0}/2$ and 
$\rho_{drip}= 3.285\times10^{11}$ g.cm$^{-3}$, extends from $R_c$=10.74~ km, which is the
radius at the core-crust interface, to 11.29~km. 

The heat equation (1) is solved in the region of the inner crust matter. Since during the
fast thermalisation the energy loss due to the neutrino emission in the crust is negligible
we take  $Q_\nu$=0. The initial temperature distribution in the crust and the boundary condition 
at the core-crust interface are chosen to simulate  a rapid cooling  process.
Thus, at $t=0$ it is supposed 
that the crust has a flat temperature, i.e., $\tilde{T}(r,t=0) =T_i$. 
We also consider that at the outer border of the crust the gradient of the temperature vanishes. 

The temperature evolution at the core-crust interface is obtained from Eq.(1) supposing that 
the radial distribution of the 
core temperature is uniform, $\partial \tilde{T}/\partial r =0$.
From Eq. (1) one thus gets
\begin{equation}
\frac{\partial T(r=R_c,t)}{\partial t} = - \frac{e^{\phi(R_c)} Q_\nu}{C_V}
\end{equation}
Considering that the neutrino emissivity in the core is given by
$Q_\nu= Q_f T_9^6$ \cite{yakhaen}, where $T_9=T/10^9$ K, and supposing that during the crust thermalisation 
the specific heat remains constant in time we finally obtain
\begin{equation}
T(r=R_c,t)=T_i\left(1+\epsilon T_i^5t\right)^{-1/5} \hbox{ with } \epsilon=\frac{5e^{\phi(R_C)}Q_f}{C_V}.
\end{equation}
For the constant $Q_f$, characterizing the fast neutrino emission in the core matter, we have taken the 
value $Q_f=10^{26}$~erg.cm$^{-3}$.s$^{-1}$  \cite{yakhaen}. 

The physical quantities which are essential for the crust thermalisation are the  thermal conductivity 
and the specific heat. In the inner crust the thermal 
conductivity is primarily determined by the electrons. Here we use the thermal 
conductivity parametrized by Lattimer et al\cite{lattimer} starting from the calculations 
of Itoh et al \cite{itoh}.  For temperatures above $10^8$ K, as used in this paper, 
the conductivity is nearly independent of the temperature and is given by
$K=C(\rho/\rho_0)^{2/3}$, where $C=10^{21}$ ergs cm$^{-1}$ s$^{-1}$ and $\rho$ is the baryonic density.

\section{ Specific heat of the inner crust matter}

The specific heat of the inner crust has contributions from the electrons, the lattice
and the unbound neutrons. They are calculated for a given set of densities in the 
Wigner-Seitz approach. Since the electrons are ultrarelativistic, they are considered 
as an uniform degenerate
gas with the specific heat given by \cite{landau}:
 \begin{equation}
 C^{(e)}_V=\frac{k_B(3\pi )^{2/3}}{3\hbar c} \left( \frac{Z}{V} \right)^{2/3}T,
\end{equation}  
where $V$ is the volume of the Wigner-Seitz cell and $Z$ is the number of the electrons
in the cell ( which is equal with the number of protons).

For the specific heat of the lattice we use the approximation employed in  Ref.\cite{pizzochero2}, 
i.e., $C_V^{lattice}=3 k_B/V$, where, as above,  $V$ is the volume of the cell and $k_B$ is the
Boltzman constant. 


We shall now discuss the temperature dependence of the specific heat of neutrons, 
which requires more elaborate calculations. The specific heat is calculated from the 
quasiparticle energies obtained solving  the HFB  equations at finite-temperature
in the Wigner-Seitz approximation. The details of the HFB calculations in 
a Wigner-Seitz cell are given in Ref. \cite{sandulescu2}.  In the HFB calculations 
the mean field is described with the same  interaction used in the
star model, i.e., the Skyrme force SLy4.  A completely consitent cooling simulation 
would require a inner crust with a Wigner-Seitz cells structure calculated from 
the same interaction. Since at present such cells structure is not available, 
here we shall use the Wigner-Seitz cells determined in Ref.\cite{negele}. 
Their properties are summarize in Table I (see the Appendix). 

The  pairing correlations in the inner crust matter are described with a  density  dependent 
contact force  of the following form \cite{bertsch}:
\begin{equation}
V (\mathbf{r}-\mathbf{r^\prime}) = V_0 [1 -\eta 
(\frac{\rho(r)}{\rho_0})^{\alpha}] 
\delta(\mathbf{r}-\mathbf{r^\prime}), 
\end{equation}
where $\rho(r)$ is the baryonic density.
To analyse  the dependence of the crust thermalisation on the intensity of pairing correlations,
in the calculations we have used two sets of parameters for the pairing force.
They are chosen to simulate two possible scenarios for pairing in neutron matter
corresponding to : (1) BCS calculations with realistic two-body interactions 
extracted from nucleon-nucleon scattering \cite{lombardo}; in this approximation the 
maximum pairing gap in uniform neutron matter is about 3 MeV; (2) calculations which go 
beyond the BCS approximation by taking into account in-medium effects on two-body 
interaction and self-energy; here we shall consider those calculations which predicts 
a maximum pairing gap in neutron matter of about 1 MeV \cite{wambach,shen}. 
These two scenarios, called below strong  and weak pairing, can be simulated by 
using two pairing forces with the same parameters for the density dependent part, 
namely $\eta=0.7$ and $\alpha=0.45$, and two different strengths, i.e, 
$V_0=\{-570,-430\}$ Mev $fm^{-3}$. These parameters have  been used with an energy
cut-off in the quasiparticle spectrum, required by the zero range of the pairing force. 
The cut-off was introduced smoothly, i.e., by an exponential factor  $e^{-E^2_i/100}$
acting for $E_i>20$ MeV, where $E_i$ are the HFB quasiparticle energies.

With the setting discussed above we have solved the HFB equations for a given 
Wigner-Seitz cell and determined the quasiparticle spectrum $E_i$ and the 
corresponding entropy, 
i.e.,
\begin{equation}
S=-k_B \sum_{i} (2j_i+1) (f_{i} \ln f_{i}+(1-f_{i})\ln (1-f_{i})).
\end{equation}
where $f_i = [1 + exp ( E_i/k_B T)]^{-1}$ is the Fermi distribution and T is the temperature.
From the entropy we then calculated the
specific heat of the neutrons
\begin{equation}
C_{V}=\frac{T}{V}\frac{\partial S}{\partial T} , 
\end{equation}
where V is the volume of the Wigner-Seitz cell.

\begin{figure}
\begin{center}
\includegraphics*[scale=0.30,angle=-90]{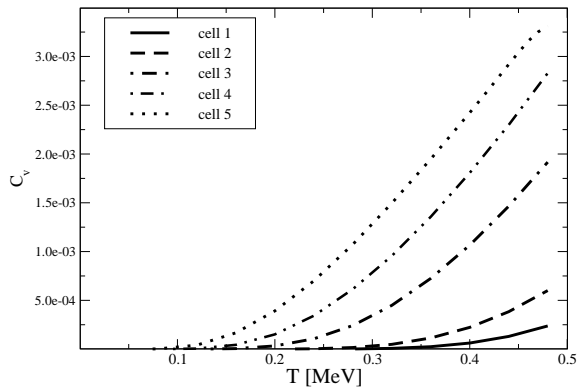}
\includegraphics*[scale=0.30,angle=-90]{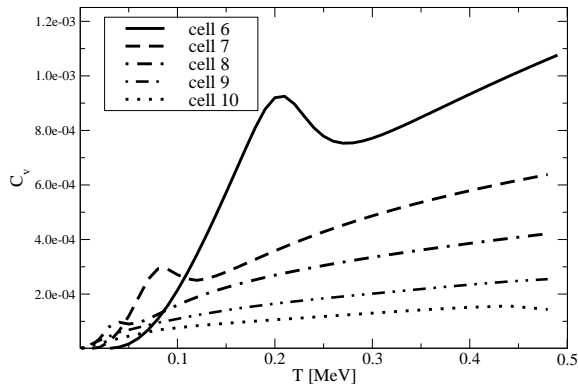}
\caption{Neutron-specific heats in various Wigner-Seitz cells for strong pairing.
The specific heat is given in units of Boltzman constant $k_B$.}
\end{center}
\end{figure}

\begin{figure}
\begin{center}
\includegraphics*[scale=0.30,angle=-90]{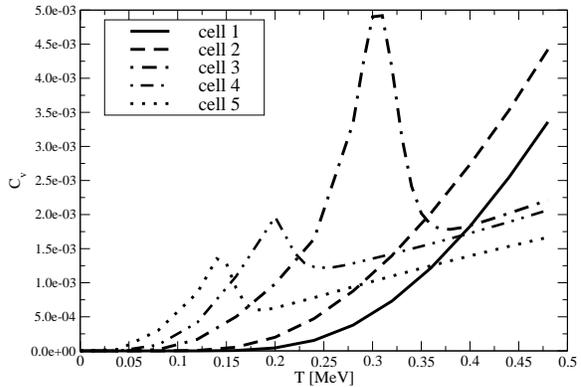}
\includegraphics*[scale=0.30,angle=-90]{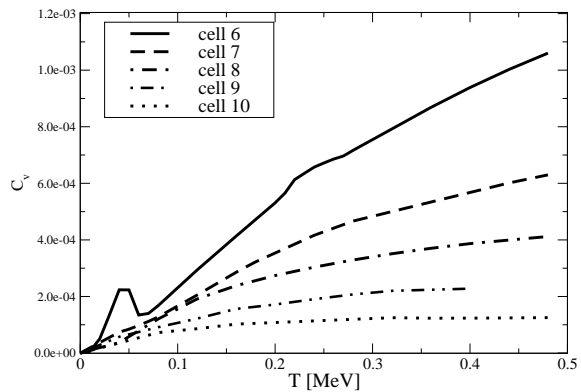}
\caption{Neutron-specific heats in various Wigner-Seitz cells for weak pairing.
The specific heat is given in units of Boltzman constant $k_B$.}
\end{center}
\end{figure}

The specific heat was calculated for all temperatures needed in the 
thermalisation process. Here we show the results for temperatures 
up to 500 keV.  How looks the temperature
dependence of specific heat for the two scenarios of the pairing intensity 
is shown in Figs.(1,2). As noticed already
in Ref.\cite{sandulescu2}, we can see that the 
specific heat has very different values for the two pairing scenarios. 
It is also interesting to notice that the
specific heats of the cells have rather different temperature dependence.
Thus, for the strong  pairing scenario, shown in Fig. 1, the specific heat 
is in the superfluid regime for the first 5 cells (upper panel).  
This is not the case for the next cells (bottom pannel) where in the same 
temperature range there is a transition from the superfluid to the normal 
phase, as clearly seen for the cells nr. 6-8 (for the last two cells the transition 
temperature cannot be noticed because is too small). On the other hand, as seen 
in Fig. 2, for the weak pairing the specific heat is entirely in the 
superfluid regime only for the first two cells. 

\begin{figure}
\begin{center}
\includegraphics*[scale=0.30,angle=-90]{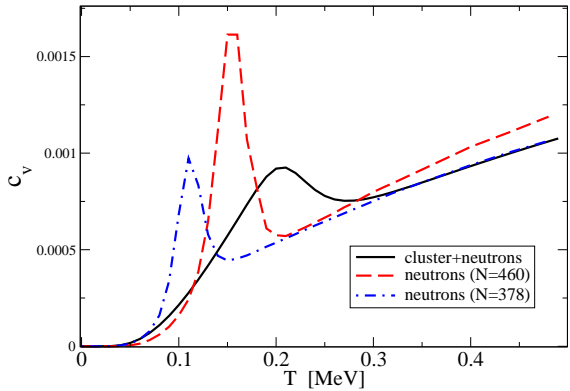}
\caption{Neutron-specific heat in the the Wigner-Seitz cell nr 6 
for strong pairing. The results corresponds to various approximations
discussed in the paper. The specific heat is given in units of Boltzman 
constant $k_B$.}
\end{center}
\end{figure}

To illustrate the particular behavior of the specific heat in non uniform matter and the
validity of various approximations, in what follows we shall discuss in more detail the 
results for the cell nr. 6, which contains N=460 neutrons and Z=40 protons (see Appendix A). 
In this cell the HFB calculations predict 378 unbound neutrons. It is interesting that 
in spite of many neutrons in the cell the number of the bound neutrons in the cluster with 
Z=40 protons is equal to the magic number 82, as for the dripline nucleus  $^{122}$Zr 
(see, e.g., Ref.\cite{122zr}). The specific heat given by the HFB 
spectrum, in which the contribution of the cluster is included, is shown in Fig.3
by full line. In the same figure are shown also the specific heats corresponding to two approximations
employed in some studies \cite{lattimer,pizzochero}. In these approximations the 
non uniform distribution of the neutrons is replaced with a uniform gas formed 
by the total number of neutrons in the cell or by taking only the number of 
the unbound neutrons. The latter case is considered as an effective way of
taking into account the influence of the cluster \cite{pizzochero}. 
How these approximations work is seen in Fig.3. To make the comparison meaningful, 
the calculations for the uniform neutron gas are done solving the HFB equations 
with the same boundary conditions as for the non uniform system, i.e., neutrons+cluster. 
As seen in Fig.3, the transition from the superfluid to the normal phase is taking
place at a lower temperature in the case of uniform neutron gas, especially when
are considered only the unbound neutrons. We can also notice that, in contrast to 
the uniform system, in the non uniform system the transition from the superfluid
to the normal phase is smooth.

\begin{figure}
\begin{center}
\includegraphics*[scale=0.30,angle=-90]{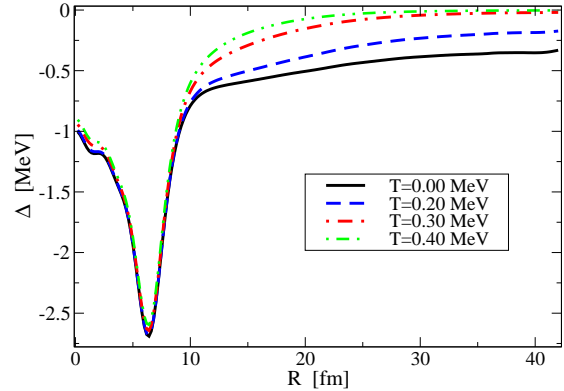}
\caption{Radial distribution of the pairing field for the Wigner-Seitz cell nr. 6 
for various temperatures. The results correspond to a strong pairing.}
\end{center}
\end{figure}

 To see better what happens in a non uniform system, in Fig.4 is shown the evolution 
with the temperature of the pairing field in the cell nr. 6. We can notice that 
at zero temperature the pairing field is much larger in the surface of the cluster
than in the bulk region. Due to this fact by increasing the temperature 
the pairs corresponding to the  neutrons localized preferentially in the 
surface region of the cluster are destroyed gradually and much slower compared 
to the pairs formed by neutrons localized far from the surface of the cluster.
In fact, as shown in Fig.5, the non uniform system shows two transition regions, 
one around T=200 keV, corresponding mainly to the neutrons located far from the 
surface of the cluster, and another one, much less pronounced, around T=900 keV,
which corresponds to the neutrons localised in the surface region of the cluster.

\begin{figure}
\begin{center}
\includegraphics*[scale=0.30,angle=-90]{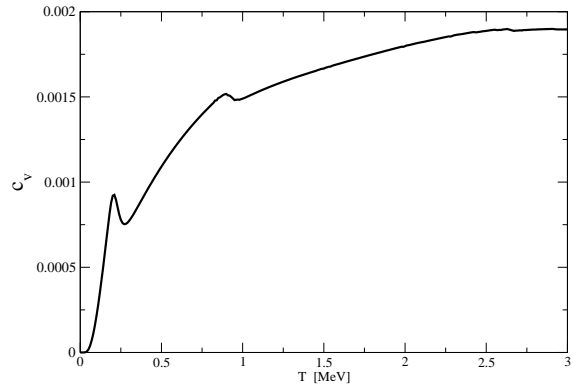}
\caption{Neutron-specific heat in the Wigner-Seitz cell nr. 6 for strong pairing. 
Here the results are shown up to high temperatures in order to illustrate the
second transition region around T= 900 keV. The specific heat is given in units of $k_B$.}
\end{center}
\end{figure}

\section{Crust thermalisation}

In this section we discuss the crust thermalisation obtained by solving the heat 
equation (1) with the specific heats presented in the previous section.
The time evolution of the temperature in the inner crust is illustrated  in 
Fig.6 which shows the results obtained for an initial temperature $T_i$=500 keV.
One can see that the crust thermalisation is strongly enhanced by the presence 
of pairing correlations.  

\begin{figure}
\begin{center}
\includegraphics*[scale=0.30,angle=0]{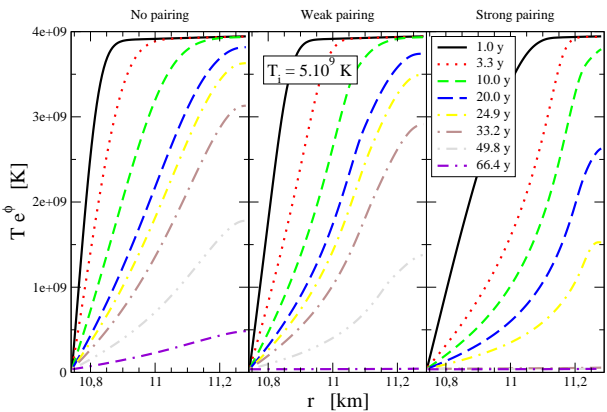}
\caption{Time evolution of the readshifted temperature $Te^{\phi}$ 
inside the inner crust for $T_i$=500 keV. The variable $r$ represents the radius of the star in $km$.}
\end{center}
\end{figure}

Of physical interest is the time evolution of the so-called apparent surface temperature
seen by a distant observer, $T_{inf}$. For a non-magnetized envelope composed by iron 
and light element the surface temperature  $T_{inf}$ is given by \cite{pochayak}
\begin{equation}
T_{inf}=T_{eff} \sqrt{1- \frac{2GM}{c^2R}}
\end{equation}
where
\begin{eqnarray}
\nonumber
T_{eff} &=& 10^6\left[ \frac{g}{10^{14}} \left((7\zeta)^{2.25}+(\zeta/3)^{1.25}\right)\right] ^{(1/4)} \\
\nonumber
\zeta &=& T_{b9}-(T_*/10^3) \\
\nonumber
g &=& GM/R\sqrt{1-GM/c^2R}
\nonumber
\end{eqnarray}
In the equations above $T_*=\left(7 T_{b9}\sqrt{g_{14}}\right)^{1/2}$ where 
$g_{14}=g/10^{14}$, $T_{b9}=T_{b}/10^9$ and $T_b$ is the temperature
at the outer edge of the inner crust provided by the heat equation.

The time evolution of the surface temperature $T_{inf}$ is displayed in Fig.7 
for three initial temperatures of the crust, $T_i=\{100,300, 500\}$ keV. 
We can clearly notice that 
the  pairing is reducing significantly 
the thermalisation time of the crust. For example, as seen  in Fig.7, bottom panel, the cooling 
time drops from about 75 years, the value obtained when the neutrons are considered in 
the normal phase, to about 55 years for weak pairing and  to about 30 years for strong pairing. 
Thus the two pairing scenarios give very  different predictions for the thermalisation time. 
The same conclusion was obtained previously in a schematic cooling model \cite{monrozeau}.
However, the latter predicts very different thermalisation times compared to the realistic
cooling model employed here.

\begin{figure}
\begin{center}
\includegraphics*[scale=0.25,angle=-90]{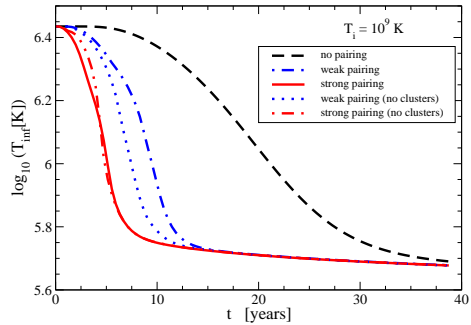}
\includegraphics*[scale=0.25,angle=-90]{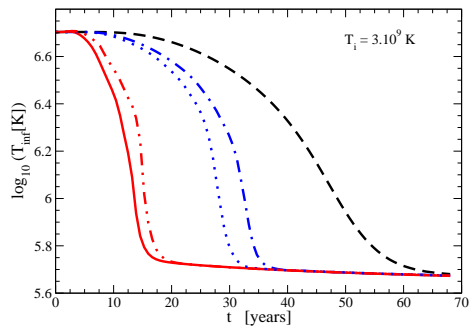}
\includegraphics*[scale=0.25,angle=-90]{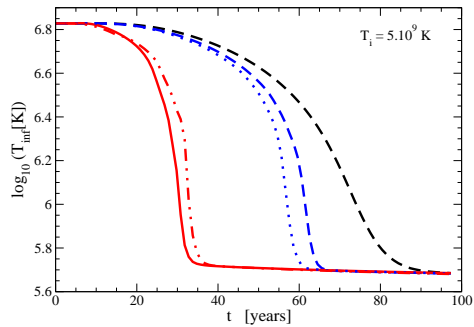}
\caption{Time evolution of the surface temperature for the 
initial temperatures $T_i=\{100,300,500\}$ keV. }
\end{center}
\end{figure}

In Fig.7 are shown also the  thermalisation times obtained by neglecting the effect
of the clusters, i.e., supposing that the baryonic matter in the inner crust
is uniform. In this case the specific heat of the neutrons is calculated
from the quasiparticle spectrum of BCS equations solved for infinite matter.
For the BCS calculations we have used the single-particle energies provided by
the asymmetric nuclear matter with the same neutron to proton fraction 
as in the Wigne-Seitz cells of Table I. As seen in Fig.7, the results for the weak 
pairing scenario indicate that the surface temperature is dropping faster for uniform 
matter than for non uniform matter. For the strong pairing the behaviour is opposite.
Thus the clusters have a non-negligible and a non-trivial effect on the time evolution 
of the surface temperature.
 
In order to test the effect of the protons on uniform matter calculations 
we have also evaluated the time evolution of the surface 
temperature considering only a uniform gas of neutrons, i.e.,
neglecting in the BCS equations the contribution of the protons to the 
single-particle energies of neutrons. The results are very similar to the
ones obtained for the uniform asymmetric matter.

\section{Summary and Conclusions}

In this paper we have studied how the thermalisation of neutron stars crust
 depends on pairing properties and the cluster structure of the inner crust matter. 
The thermal evolution was obtained by solving the 
relativistic heat equation with initial conditions specific to a rapid 
cooling process and supposing  that during the thermalisation 
there are no sinks or sources of energy in the crust. The specific heat
of neutrons was  calculated from the HFB spectrum, taking into accound 
consistently the effects of nuclear clusters, pairing correlations and 
temperature. The thermal evolution of the inner crust was analysed using 
for the neutrons two sets  of specific heats obtained with a strong and 
a weak pairing force which simulate two possible scenarios for the 
intensity of pairing correlations in neutron matter. The results
show that the crust thermalisation is strongly influenced
by the intensity of pairing correlation. This result confirm what it
was found earlier in Ref. \cite{monrozeau} with a schematic cooling model.
However, the latter predicts thermalisation times which are very
different from the results obtained with the realistic cooling model
employed in this study. We have also shown that the cluster structure 
of the inner crust affects significantly and in a non-trivial way the  
time evolution of the surface temperature, mainly for weak pairing and
before the thermal equilibrium between the crust and the core is reached.

\vskip 0.2cm
{\bf Acknowledgment.} We thank P. M. Pizzochero for valuable discussions and for
his help in solving the heat equation.
This work was supported by ESF through the project " The New Physics of Compact Stars"
and by CNCSIS through the grant IDEI no. 270.

\section {Appendix A}

In this appendix are summarized the properties of the Wigner-Seitz
cells determined in Ref. \cite{negele} and used in this paper.
Compared to Ref. \cite{negele}, here we didn't include the most
dense cell, which is close to the region of more complicated pasta
phases. 
\begin{table}[htb]
\centering
\begin{tabular}{c c c c c c}
\hbox {cell} & N &Z &$R_{WS}$&$\rho$ \\
 & & &$[fm]$& $[g.cm^{-3}]$ \\
\hline
\hline
10 & 140 &40  &54 &$4.7\times 10^{11}$ \\
 9 & 160 &40  &49 &$6.7\times 10^{11}$ \\
 8 & 210 &40  &46 &$1.0\times 10^{12}$ \\
 7 & 280 &40  &44 &$1.5\times 10^{12}$ \\
 6 & 460 &40  &42 &$2.7\times 10^{12}$  \\
 5 & 900 &50  &39 &$6.2\times 10^{12}$  \\
 4 &1050 &50  &36 &$9.7\times 10^{12}$ \\
 3 &1300 &50  &33 &$1.5\times 10^{13}$ \\
 2 &1750 &50  &28 &$3.3\times 10^{13}$ \\
 1 &1460 &40  &20 &$7.8\times 10^{13}$ \\
\end{tabular}
\caption{The structure of the cells determind in Ref. \cite{negele} , i.e., 
the baryonic densities ($\rho$), the number of neutrons (N), the number 
of protons (Z) and the cell 
radii ($R_{WS}$)}
\label{tab3}
\end{table}


\end{document}